\documentclass{article}
\usepackage{graphicx}

\def\l{\lambda}

\def\ds{\displaystyle}
\def\be{\begin{equation}}
\def\ee{\end{equation}}
\def\beq{\begin{eqnarray}}
\def\eeq{\end{eqnarray}}

\begin{document}
\title{\bf 
Quantum Correlations and Number Theory}

\author{
     H. E. Boos \\
{\it Institute for High Energy Physics}\\
{\it Protvino, 142284, Russia}\\
\\     V. E.  Korepin \\
{\it C.N.~Yang Institute for Theoretical Physics}\\
{\it State University of New York at Stony Brook}\\
{\it Stony Brook, NY 11794--3840, USA}\\
\\     Y. Nishiyama \\
{\it Department of Physics, Faculty of Science, Okayama University}\\
{\it Okayama 700-8530, Japan}\\
\\     M. Shiroishi \\ 
{\it Institute for Solid State Physics, University of Tokyo}\\
{\it Kashiwanoha 5-1-5, Kashiwa, Chiba, 277-8571, Japan}
}

\maketitle\thispagestyle{empty}
\abstract{ We study spin-1/2 Heisenberg ${XXX}$  antiferromagnet.
The spectrum of the Hamiltonian was found by Hans Bethe in 1931, \cite{Bethe}.
We study the probability of  formation of ferromagnetic string in the antiferromagnetic ground state, 
which we call emptiness formation probability ${P(n)}$. This is the most fundamental correlation function.  We prove that for the
 short strings it can be expressed in terms of the Riemann zeta function with odd arguments, logarithm ${\ln 2}$ and rational coefficients. 
This adds yet another link between statistical mechanics and number theory.
We have obtained an analytical formula for ${P(5)}$ for the first time.  We have also calculated ${P(n)}$
 numerically by the Density Matrix Renormalization Group. The results agree quite well with the analytical ones.
 Furthermore we study asymptotic behavior of   ${P(n)}$ at finite  temperature by Quantum Monte-Carlo
 simulation. It also agrees with our previous analytical results.
 }
\section{Introduction}
\setcounter{equation}{0}
\renewcommand{\theequation}{1.\arabic{equation}}
Recently the considerable progress has been achieved for the exact calculation of correlation
 functions in the spin-1/2 
Heisenberg ${XXZ}$  chain \cite{Korepin93, Jimbo95, Korepin94, Essler95n2, Essler95, Jimbo96, Kitanine00, Razumov00, Boos01, Boos02,
Shiroishi01}. Hans Bethe discovered his Ansatz, while diagonalising  ${XXX}$ Hamiltonian.
The most important features of exactly solvable models, like two-body reducibility of dynamics
were first discovered for this model.
  We believe that  the antiferromagnetic Heisenberg ${XXX}$ chain is one of the most fundamental exactly solvable models.
%we expect especially simple results.
 Recently we  developed a new method of evaluation of  the multi-integral
 representation of correlation functions \cite{Boos01,Boos02}. We study the most fundamental correlation 
function of the model, which we
 call {\it emptiness formation probability}  ${P(n)}$. We shall abbreviate it to EFP.
 It was first  introduced in 
\cite{Korepin94}
\begin{equation}
P(n) = \langle {\rm GS} | \prod^n_{j=1} P_j | {\rm GS} \rangle,
\end{equation}
where ${P_j = S^z_j+{1\over2}}$ is the projector on the state with the spin up in the ${j}$-th lattice site. ${| {\rm GS} \rangle}$ is 
the antiferromagnetic ground state in the thermodynamic limit constructed by Hulth\'{e}n \cite{Hulthen38}. ${P(n)}$ is a probability of formation of 
a ferromagnetic string of length ${n}$ in ${| {\rm GS} \rangle}$.

The Hamiltonian of the  Heisenberg ${XXX}$ chain  is given by
\begin{equation}
H= J \sum^N_{j=1} \left( S^x_jS^x_{j+1}+S^y_jS^y_{j+1}+ S^z_jS^z_{j+1} 
 - \frac{1}{4} \right), \label{XXX}
\end{equation}
where the coupling constant ${J}$ is positive for the antiferromagnet.  

In ref. \cite{Boos01,Boos02}, we wrote the Hamiltonian (\ref{XXX}) in terms of the Pauli matrices. It  corresponds to  ${J=4}$. 
Note that the Hamiltonian (\ref{XXX}) annihilates the ferromagnetic states ${| \uparrow \dots \uparrow \rangle }$ 
(or ${| \downarrow \dots \downarrow \rangle}$).

First cases ${P(3)}$ and ${P(4)}$ were 
 calculated  by  means of  the multi-integral representation in \cite{Boos01,Boos02}. 
%The obtained formula for ${P(3)}$ coincides with the known one.
% The formula for ${P(4)}$ was  new.
 In this paper, 
we present a new analytic formula for  ${P(5)}$ 

\begin{eqnarray}
P(5) &=& \frac{1}{6} - \frac{10}{3} \ln 2  + \frac{281}{24} \zeta(3) - \frac{45}{2} \ln 2 \cdot \zeta(3) - \frac{489}{16} \zeta(3)^2 \nonumber \\
& & \ \ \ \  - \frac{6775}{192} \zeta(5) + \frac{1225}{6} \ln 2 \cdot \zeta(5) - \frac{425}{64} \zeta(3) \cdot \zeta(5) - 
\frac{12125}{256} \zeta(5)^2 \nonumber \\
& & \ \ \ \ + \frac{6223}{256} \zeta(7) - \frac{11515}{64} \ln 2 \cdot \zeta(7) + \frac{42777}{512} \zeta(3) \cdot \zeta(7) \nonumber  \\
&=& 2.011725953 \times 10^{-6}, \label{P5}
\end{eqnarray}
where ${\zeta(s)}$ is the Riemann zeta function
\begin{equation}
\zeta(s) = \sum_{n=1}^{\infty} \frac{1}{n^s}, \ \ \ \ {\rm for \ } \Re(s) >1.  
\end{equation}

 ${P(5)}$ is expressed in terms of ${\ln 2, \zeta(3), \zeta(5)}$ and ${\zeta(7)}$ with rational coefficients. 
In fact, it was conjectured \cite{Boos02}: \\
 ``{\bf ${P(n)}$ is always expressed in terms of logarithm ${\ln 2}$ ,  Riemann zeta functions ${\zeta(2k+1)}$ 
with odd argument
 and  rational coefficients.}''  \\

This means that all values of  ${P(n)}$  are different transcendental numbers  \cite{tr} .

For comparison, let us list  previous results for ${P(n)}$.
\begin{eqnarray}
P(1) &=& \frac{1}{2} = 0.5, \label{P1} \\
P(2) &=& \frac{1}{3} - \frac{1}{3} \ln 2 = 0.102284273, \label{P2} \\
P(3) &=& \frac{1}{4} - \ln 2 + \frac{3}{8} \zeta(3) = 0.007624158, \label{P3} \\
P(4) &=& \frac{1}{5} - 2 \ln 2 + \frac{173}{60} \zeta(3) - \frac{11}{6} \ln 2 \cdot \zeta(3) - \frac{51}{80} \zeta^2(3)  \nonumber \\
& & \ \ \ \ - \frac{55}{24} \zeta(5) + \frac{85}{24} \ln 2 \cdot \zeta(5) = 0.000206270. \label{P4}
\end{eqnarray}
Let us mention that in contrast with  $P(1), P(2), P(3)$ [ which do not contain non-linear terms of the
 Riemann zeta function]
the values $P(4)$ and $P(5)$ do contain non-linear terms in  Riemann zeta function.
It might be instructive to express formulae (\ref{P5}),\ (\ref{P4}) for $P(5)$ and $P(4)$ in a linear form
 by introducing the multiple zeta values \cite{Zagier} 
\be
\zeta(k_1,k_2,\ldots ,k_m)\;=\;\sum_{n_1>n_2>\ldots >n_m>0} 
n_1^{-k_1} n_2^{-k_2}\ldots n_m^{-k_m}.
\label{mzv}
\ee
The length (or depth) of this  multiple zeta value is equal to  $m$ and the level (or weight) is equal to 
$k_1\, + \, k_2\, +\,\ldots \,+\, k_m$.
The result looks as follows:
\begin{eqnarray}
P(4) &=& \frac{1}{5} - 2 \ln 2 + \frac{173}{60} \zeta(3) - \frac{11}{6} \ln 2 \cdot \zeta(3) - \frac{55}{24}\zeta(5)+
\frac{85}{24}\ln 2 \cdot \zeta(5)  \nonumber \\
& & \ \ \ \ - \frac{51}{10} \zeta(3,3) + \frac{153}{80} \zeta(4,2), 
\label{PP4} \\
P(5) &=& \frac{1}{6} - \frac{10}{3} \ln 2  + \frac{281}{24} \zeta(3) - \frac{45}{2} \ln 2 \cdot \zeta(3) - \frac{6775}{192} \zeta(5) \nonumber \\
& & \ \ \ \   + \frac{1225}{6} \ln 2 \cdot \zeta(5) + \frac{6223}{256} \zeta(7) - \frac{11515}{64} \ln 2 \cdot \zeta(7)  
\nonumber \\
& & \ \ \ \  -\frac{489}{2} \zeta(3,3) + \frac{1467}{16}\zeta(4,2)
-\frac{12495}{128} \zeta(4,4) - \frac{85}{64}\zeta(5,3) -\frac{425}{128}\zeta(6,2)\nonumber\\
& &\ \ \ \
+\frac{487037}{512} \zeta(5,5) +
\frac{584425}{1024} \zeta(6,4) + \frac{596647}{1024}\zeta(7,3) +\frac{42777}{1024}\zeta(8,2).
\label{PP5}
\end{eqnarray}
Here  we  used the following identities
\beq
\zeta(3)^2 &=& 8\zeta(3,3)\,-\,3\zeta(4,2),\nonumber\\
\zeta(3) \cdot \zeta(5) &=& \frac{147}{10}\zeta(4,4) + \frac{1}{5}\zeta(5,3) + \frac{1}{2} \zeta(6,2), \nonumber\\
\zeta(5)^2 &=& 22\zeta(5,5) + 10\zeta(6,4) + 10\zeta(7,3),\nonumber\\
\zeta(3) \cdot \zeta(7) &=& \frac{167}{7}\zeta(5,5) + \frac{25}{2}\zeta(6,4) + \frac{177}{14} \zeta(7,3)
+\frac{1}{2}\zeta(8,2).
\label{equal}
\eeq
These identities can be derived using the reccurent relations (4.1), (4.5),..., (4.7) of ref. \cite{MSch}. 
Let us note that the Drinfeld's associator is also related to multiple zeta values in a linear way \cite{Le-Mur}.

Since, $\ln 2$ in formulae above looks somehow isolated, we believe that it seems to be more appropriate to 
express $P(n)$ in terms of the alternating zeta series (the value of polylogarithm at root of unity)
\be
\zeta_a(s)\;=\;\sum_{n>0}{(-1)^{n-1}\over n^s}\;=\; - \mbox{Li}_s(-1)
\label{za}
\ee
Here $\mbox{Li}_s(x)$ is the polylogarithm. 
The alternating zeta series is related to the Riemann zeta function as follows
\be
\zeta(s)\;=\;{1\over 1-2^{1-s}}\zeta_a(s)
\label{za1}
\ee
This formula is valid for $s\neq 1$.
In contrast with the zeta function [which has the pole when $s\rightarrow 1$], the alternating zeta has a limit 
as  $s\rightarrow 1$
% does not have such a pole
%but has a good limit when $s\rightarrow 1$, namely,
\be
\zeta_a(1)\;=\;\ln 2
\label{zaat1}
\ee
Using the formulae (\ref{P5}),...,(\ref{P4}), (\ref{za1}) and (\ref{zaat1}), one can get the five first values of $P(n)$ expressed
via the alternating zeta series
\beq
P(1) &=& \frac{1}{2}, \nonumber \\
P(2) &=& \frac{1}{3} \bigl\{ 1 - \zeta_a(1) \bigr\}, \nonumber \\
P(3) &=& \frac{1}{4} \bigl\{ 1 - 4 \zeta_a(1) + 2\zeta_a(3) \bigr\}, \nonumber \\
P(4) &=& \frac{1}{5} \bigl\{ 1 - 10 \zeta_a(1) + \frac{173}{9}\zeta_a(3) - \frac{110}{9} \zeta_a(5) - \frac{110}{9} \zeta_a(1) \cdot \zeta_a(3) \nonumber \\
& & \ \ \ \ + \frac{170}{9} \zeta_a(1) \cdot \zeta_a(5) - \frac{17}{3} \zeta_a^2(3) 
\bigr\}, \nonumber \\
P(5) &=& \frac{1}{6} \big\{ 1 - 20 \zeta_a(1) + \frac{281}{3} \zeta_a(3) - \frac{1355}{6} \zeta_a(5) + \frac{889}{6} \zeta_a(7) \nonumber \\
& & \ \ \ \ - 180 \zeta_a(1) \cdot \zeta_a(3) +  \frac{3920}{3} \zeta_a(1) \cdot \zeta_a(5)- 
\frac{3290}{3} \zeta_a(1) \cdot \zeta_a(7) - \frac{170}{3} \zeta_a(3) \cdot \zeta_a(5) \nonumber \\
& & \ \ \ \ + 679 \zeta_a(3) \cdot \zeta_a(7) - 326 \zeta_a^2(3) - \frac{970}{3} \zeta_a^2(5) \bigr\}.
\nonumber \\
\label{P1-P5a} 
\eeq
 
The values of the Riemann zeta function at odd arguments appear in several places in theoretical physics,
 not to mention  in pure mathematics. 
 Transcendental number ${\zeta(3)}$  first appeared in the expression for
correlation functions  in Takahashi's papers \cite{Takahashi77,Takahashi99}.
He evaluated the  second neighbor correlation 
%\cite{Takahashi77,Takahashi99}
\begin{equation}
\langle {\bf S}_i \cdot {\bf S}_{i+2} \rangle
 = \frac{1}{4} - 4 \ln 2 + \frac{9}{4} \zeta(3) \label{nnc}
\end{equation}
It was obtained from the ${1/U}$ expansion of the ground state energy for the half-filled Hubbard chain 
(see also another derivation in ref. \cite{Dittrich97}).  We remark that the expression for ${P(3)}$ (\ref{P3}) can be extracted from (\ref{nnc}).

Now let us discuss the asymptotic behavior of ${P(n)}$ when ${n}$ is large. At zero temperature we 
believe that ${P(n)}$ 
%for the ${XXX}$ Heisenberg chain
 should show a Gaussian decay as ${n}$ tends to infinity \cite{Boos01,Boos02}.
  In order 
to prove it mathematically we have to obtain a general formula for ${P(n)}$ which has not been achieved yet.
 On the other hand, 
we can calculate ${P(n)}$ by numerical means and confirm our analytical results. We have applied the Density Matrix 
Renormalization Group (DMRG) method 
\cite{White92,White93,Peschel99}
and obtained  more numerical values of ${P(n)}$. The result is given in the end of  section 2.  At finite temperature it was shown that 
${P(n)}$ decays exponentially \cite{Boos01}. This time we can employ the Quantum Monte-Carlo (QMC) simulation \cite{Suzuki76} to calculate 
${P(n)}$ numerically. In section 3 we confirm that ${P(n)}$ exhibits an exponential decay at finite temperature
and confirm again our  analytical expression. 
Let us note that this numerical approach was successfully applied to the ${XX}$ model in ref. \cite{Shiroishi01}.

\section{${P(n)}$ at zero temperature}
\setcounter{equation}{0}
\renewcommand{\theequation}{2.\arabic{equation}}
The integral representation of ${P(n)}$ for the ${XXX}$ chain was obtained 
in \cite{Korepin94} based on the vertex operator approach \cite{Jimbo95}
\begin{equation}
P(n)=\int_C \frac{d \lambda_1}{2 \pi i \lambda_1}
\int_C \frac{d \lambda_2}{2 \pi i\lambda_2} \ldots
\int_C \frac{d \lambda_n}{2 \pi i\lambda_n}
\prod_{a=1}^n (1+\frac{i}{\lambda_a})^{n-a}
(\frac{\pi \lambda_a}{\sinh \pi \lambda_a})^n
\prod_{1 \le j<k \le n} 
\frac{\sinh \pi (\lambda_k-\lambda_j)}{\pi (\lambda_k-\lambda_j-i)}. \nonumber \\
\label{intPn} 
\end{equation}
The contour $C$ in each integral goes parallel to the real axis
with the imaginary part between $0$ and $-i$.
Below we sketch the  evaluation of  the integral  for ${P(5)}$. It can be written in the following form
\begin{equation}
P(5)=
\prod_{j=1}^5 \int_{C} \frac{d\lambda_j}{2\pi i }
U_5(\lambda_1,\ldots,\lambda_5) T_5(\lambda_1,\ldots,\lambda_5),
\label{intPna}
\end{equation}
where
\begin{equation}
U_5(\lambda_1,\ldots,\lambda_5) = \pi^{15} 
\frac{\prod_{1 \leq k < j \leq 5} \sinh{\pi(\lambda_j-\lambda_k)}}{
\prod_{j=1}^5 \sinh^5 {\pi \lambda_j}
},
\label{U_5}
\end{equation}
and
\begin{equation}
T_5(\lambda_1,\ldots,\lambda_5) =
\frac{\prod_{j=1}^5 \lambda_j^{j-1}(\lambda_j+i)^{5-j}}{
\prod_{1 \leq k < j\leq 5}(\lambda_j-\lambda_k-i)
}.
\label{T_5}
\end{equation}

Taking into account the properties of the functions ${U_5(\lambda_1,\ldots,\lambda_5)}$, one can reduce the integrand ${T_5(\lambda_1,\ldots,\lambda_5)}$  to the ``canonical form",  ${T^c_5(\lambda_1,\ldots,\lambda_5)}$ 
as was done for $P(2)$, ${P(3)}$ and ${P(4)}$ in ref. \cite{Boos02}
\begin{equation}
T^c_5(\lambda_1,\ldots,\lambda_5) =  P_0^{(5)}\,+\,{P_1^{(5)}\over\l_2-\l_1}\,+
{P_2^{(5)}\over (\l_2-\l_1)(\l_4-\l_3)},
\label{Tc5}
\end{equation}
where the $P^{(0)},P^{(1)}, P^{(2)}$ are polynomials of the integration variables $\l_1,\ldots,\l_5$.
The manifest form of these polynomials is shown in Appendix.

Now using the methods developed in \cite{Boos02} we can calculate three integrals that contribute into $P(5)$
\be
J_0^{(5)}=
\prod_{j=1}^5 \int_{C} \frac{d\lambda_j}{2\pi i }
U_5(\lambda_1,\ldots,\lambda_5) P_0^{(5)}(\lambda_1,\ldots,\lambda_5),
\label{J0}
\end{equation}
\be
J_1^{(5)}=
\prod_{j=1}^5 \int_{C} \frac{d\lambda_j}{2\pi i }
U_5(\lambda_1,\ldots,\lambda_5) 
{P_1^{(5)}(\lambda_1,\ldots,\lambda_5)\over \l_2-\l_1},
\label{J1}
\ee
\be
J_2^{(5)}=
\prod_{j=1}^5 \int_{C} \frac{d\lambda_j}{2\pi i }
U_5(\lambda_1,\ldots,\lambda_5) 
{P_2^{(5)}(\lambda_1,\ldots,\lambda_5)\over (\l_2-\l_1)(\l_4-\l_3)}.
\label{J2}
\end{equation}
The result looks as follows:
\beq
J_0^{(5)} &=& \frac{689}{576},
\label{J0res} \\ 
J_1^{(5)} &=& -\frac{593}{576} - \frac{10}{3} \ln{2} - \frac{2773}{384} \zeta(3) - \frac{175}{48} \zeta(5) + \frac{13727}{1024} \zeta(7),
\label{J1res} \\
J_2^{(5)} &=& \frac{2423}{128} \zeta(3) - \frac{45}{2} \ln{2} \cdot \zeta(3)- \frac{489}{16}\zeta(3)^2 - \frac{2025}{64} \zeta(5) \nonumber\\
& & \ \ + \frac{1225}{6} \ln{2} \cdot \zeta(5) - \frac{425}{64} \zeta(3) \cdot \zeta(5) - \frac{12125}{256} \zeta(5)^2 \nonumber\\ 
& & \ \ + \frac{11165}{1024} \zeta(7) - \frac{11515}{64}\ln{2} \cdot \zeta(7) + \frac{42777}{512} \zeta(3) \cdot \zeta(7). 
\label{J2res}
\eeq
Summing up these three values we come to our final answer (\ref{P5}).

Using the same method we can, in principle, get the analytic formula for any ${P(n)}$.  Unfortunately, so far we have not succeeded in calculating 
$P(n)$ for $  n \ge 6$.
On the other hand, one can estimate the numerical values of ${P(n)}$ using DMRG \cite{White92,White93}.  
The method is suitable for studying ground-state properties. We followed standard algorithm, which can be found
in literature (see ref. \cite{Peschel99}).
Below we shall outline some technical points that are
relevant for the simulation precision.
We implemented the infinite-system method. We have repeated 
renormalization 200-times.
At each renormalization, we kept, at most, 200 relevant states
for a (new) block; namely, we set $m=200$.
The density-matrix eigenvalue $\{ w_\alpha \}$
of remaining bases indicates the statistical weight.
We found ${w_\alpha>10^{-10}}$ : 
That is, we have remained almost all relevant states 
with appreciable statistical weight $w_\alpha>10^{-10}$
through numerical renormalizations.
In other words, we have discarded (disregarded) those
states with exceedingly small statistical weight $w_\alpha<10^{-10}$,
which may indicate error of the present simulation.

The obtained data are shown in Table \ref{table1}.

\begin{table}[htbp]
\begin{center}
\caption{DMRG data for ${P(n)}$  with uncertainties in the final digits.}
\label{table1}
\begin{tabular}{@{\hspace{\tabcolsep}\extracolsep{\fill}}rrr} \hline
    ${n}$ & ${P(n)}$ by DMRG    \\ \hline
2     &  ${1.022\underline{2} \times 10^{-1} \ }$  \\ 
3     &  ${7.623\underline{8} \times 10^{-3} \ }$  \\ 
4     &  ${2.06\underline{0} \times 10^{-4} \ }$  \\ 
5     &  ${2.01\underline{0} \times 10^{-6} \ }$  \\ 
6     &  ${7.0\underline{5} \times 10^{-9} \ }$  \\ 
7     &  ${8.8\underline{5} \times 10^{-12}}$  \\ 
8     &  ${3.\underline{7} \times 10^{-15}}$ \\ \hline
\end{tabular}
\end{center}
\end{table}

Compared with the exact values (\ref{P5}), (\ref{P1}),...,(\ref{P4}), we see the DMRG data achieve about 3 digits accuracy up to ${P(5)}$. 
Probably the other data, i.e.,${P(6),...,P(8)}$ also maintain at least 1 or 2 digits accuracy.  Then combining the exact values up to ${P(5)}$ 
and the numerical data in Table 1, we have made a semi-log plot in Fig. 1. On the vertical axis we plotted the $\ln P(n)$, on
horizontal axis we plotted $n^2$.  From Fig.1 
we clearly see the data fall into a straight line suggesting that the asymptotic form of ${P(n)}$ is governed by the Gaussian form, 
\begin{equation}
P(n)  \sim a^{- n^2}.
\end{equation}
We can read off the Gaussian decay rate $a$ from the slope.
Our estimate is ${a=1.6719 \pm 0.0005}$. It is an intriguing open problem to 
associate this number with a certain analytical expression.  Doctor A.G. Abanov confirmed Gaussian form
of asymptotic expression in the frame of bosonization technique.

\begin{figure}[htbp]
\begin{center} 
\includegraphics{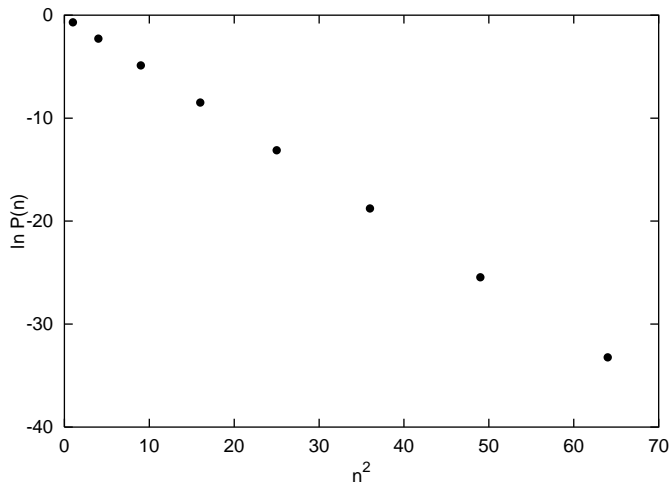}
\end{center}
\caption{$P(n)$ at zero temperature}
\label{DMRG} 
\end{figure}
\section{${P(n)}$ at finite temperature by QMC simulation}
\setcounter{equation}{0}
\renewcommand{\theequation}{3.\arabic{equation}}
At finite temperature ${T}$ the emptiness formation probability ${P(n)}$ is defined by the thermal average 

\begin{equation}
P(n) = \frac{{\rm Tr \ } \left\{ {\rm e}^{-H/T} \prod_{j=1}^{n} P_j \right\}}{{\rm Tr \ } \left\{ {\rm e}^{-H/T} \right\} }.
\end{equation}

It was shown in ref. \cite{Boos01} that ${P(n)}$ decays exponentially at finite temperature as  ${n}$ goes to infinity 
\begin{equation}
P(n) \sim c_0(T)  {\rm e}^{\frac{n f}{T}}.  \label{asPnT}
\end{equation}
Here ${f}$ is the free energy per site of the system and ${c_0(T)}$ is a constant prefactor. 
We shall confirm this formula by calculating ${P(n)}$ numerically by QMC simulation \cite{Suzuki76}. 
We adopted the continuous-time algorithm \cite{Beard96} with the cluster-flip update \cite{Evertz93,Wiese94,Kawashima94} which is completely free from the Trotter-decomposition error.
We treated system sizes up to $N=128$, and imposed the periodic boundary
condition.
We performed five-million Monte-Carlo steps initiated by 0.5 million
steps for reaching thermal equilibrium.
$P(n)$ is measured over the five-million Monte-Carlo steps. The semi-log plots of the data for several different temperatures  
are shown in Fig. 2. The straight lines (slopes) are the analytic asymptotic formula (\ref{asPnT}).
 We assumed ${c_0(T)}=1$  for simplicity. 
Actually ${c_0(T)}$ deviates from  unity  for low temperature. Anyway we observe that
 as ${n \rightarrow \infty}$,  EFP ${P(n)}$ 
decays exponentially  according to the  asymptotic form (\ref{asPnT}). In particular, at
 sufficiently
 high temperatures, even  for small ${n}$,  EFP
${P(n)}$
is well fitted by (\ref{asPnT}). In contrast, at low temperature , when ${n}$ becomes small $P(n)$ may  reflect the Gaussian decay at zero 
temperature and deviates from (\ref{asPnT}). It agrees with the physical picture. Since, zero
 temperature is a critical 
point, we expect qualitative change of asymptotic of correlation functions.

Finally let us make a comment how we can evaluate the free energy per site ${f}$ from the point of view of the Bethe Ansatz. 
There are now three different integral equations which determine 
the free energy ${f}$; 
\begin{enumerate}
\item Thermodynamic Bethe Ansatz (TBA) equations formulated by Takahashi \cite{Takahashi71} based on the string hypothesis. 
\item Non-linear Integral Equations (NLIE) found by Kl\"{u}mper \cite{Kluemper92,Kluemper93} and Destri--de Vega \cite{Destri92} 
in the development of the quantum transfer matrix method \cite{Suzuki87,Koma87,JSuzuki90,Takahashi91n1}.
\item New integral equation derived by Takahashi \cite{Takahashi00}.  
\end{enumerate}
The third one was recently discovered in an attempt to simplify the TBA equations \cite{Takahashi00}. It has also a close connection with 
the quantum transfer matrix \cite{Takahashi01}. The equation is explicitly given by
\begin{eqnarray}
u(x)&=& 2 + \oint_C \Bigg( \frac{1}{x - y - 2 {\rm i}} \exp \left[\frac{2 J/T}{(y + {\rm i})^2+1} \right] + \frac{1}{x - y + 2 {\rm i}} 
\exp \left[\frac{2 J/T}{(y - {\rm i})^2+1} \right] \Bigg) \frac{1}{u(y)} \frac{{\rm d} y}{2 \pi \rm i}, \nonumber \\
f &=& - T \ln u(0),
\label{TakahashiEq} 
\end{eqnarray}
where the contour ${C}$ is a loop which counterclockwise encircles the origin.
Numerically these three approaches provide perfectly the same data for free energy per site ${f}$.

\begin{figure}[htbp]
\begin{center} 
\includegraphics{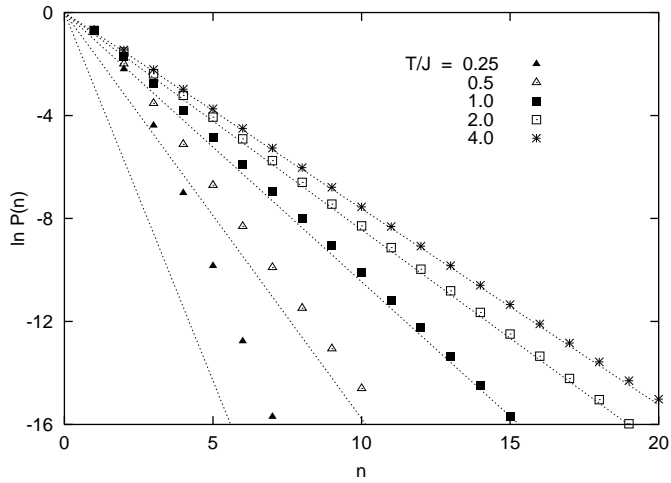}
\end{center}
\caption{$P(n)$ at finite temperature}
\label{QMC} 
\end{figure}

\section{Conclusion}
\setcounter{equation}{0}
\renewcommand{\theequation}{4.\arabic{equation}}

Let us mention again that the main result of this paper is the calculation of ${P(5)}$ by means of the multi-integral representation. 
It is expressed only 
in terms of ${\ln 2}$ and Riemann zeta function with odd arguments with rational coefficients. 
This should be the general  property of ${P(n)}$. We expect that other correlation functions such 
as ${\langle {\bf S}_j {\bf S}_{k}  \rangle}$ will share this property. 

We have calculated  numerically the value of ${P(n)}$ and considered the asymptotic  behavior. 
As was discussed in ref. \cite{Boos01}, EFP  ${P(n)}$ shows a Gaussian decay at zero temperature, 
while it decays exponentially at finite temperature. \\

\vspace{5mm}

%%%%%%%%%%%%%%%%%%%%%%%%%%%%%%%%%%%%%%%%%%%%%%%%%%%%%%%%%%%%

\begin{center}

{\bf Acknowledgment}

\end{center}

The authors are grateful to A.G. Abanov, G.P.~Pronko, A.V.~Razumov, A.P.~Samokhin, M.A.~Semenov-Tian-Shansky, M.~Takahashi and V.N.~Tolstoy 
for stimulating discussions. This research was supported by the following grants:
NSF PHY-9988566, the Russian Foundation for Basic Research under grant 
\# 01--01--00201, INTAS under grant \# 00--00561, MEXT under Grant-in-Aid for Scientific Research, \# 11440103 and \# 13740240.  

\appendix
\section{Appendix}
\setcounter{equation}{0}
\renewcommand{\theequation}{A.\arabic{equation}}
Here we show the polynomials $P_0^{(5)}, P_1^{(5)}, P_2^{(5)}$ which participate in the ``canonical form'' $T_5^c$ given by
the formula (\ref{Tc5})

\begin{equation}
P_0^{(5)} = {689\over 18} \l_2\l_3^2\l_4^3\l_5^4
\label{P05}
\end{equation}
\beq
P_1^{(5)} \quad =  
&\ds
{796333\over 15120}\l_4\l_5^2 - 
{3299\, i\over 63}\l_1\l_4\l_5^2 +
{13844\over 315} \l_1^2\l_4\l_5^2 - 
{16517\, i\over 105}\l_1^3\l_4\l_5^2 - 
{13463\, i\over 1512}\l_4\l_5^3 + \quad\quad\quad &
\nonumber\\
&\ds
{1217491\over 7560}\l_1\l_4\l_5^3 - 
{328189\, i\over 840}\l_1^2\l_4\l_5^3 - 
{279917\over 945} \l_1^3 \l_4 \l_5^3 +
{57619\over 630}\l_4^2\l_5^3 - 
{543079\, i\over 1260}\l_1\l_4^2\l_5^3 -&
\nonumber\\
&\ds
\quad\quad {9635\over 14}\l_1^2\l_4^2\l_5^3 + 
{36199\, i\over 126}\l_1^3\l_4^2\l_5^3 - 
{811901\, i\over 7560}\l_3\l_4^2\l_5^3 - 
{392107\over 1260}\l_1\l_3\l_4^2\l_5^3 +
{12503\, i\over 36}\l_1^2\l_3\l_4^2\l_5^3 +&
\nonumber\\
&\ds
{8\over 7}\l_1^3\l_3\l_4^2\l_5^3 - 
{197459\over 3780}\l_4\l_5^4 - 
{39169\, i\over 1512}\l_1\l_4\l_5^4 - 
{20873\over 180}\l_1^2\l_4\l_5^4 + 
{66119\, i\over 756}\l_1^3\l_4\l_5^4 + \quad\quad  &
\nonumber\\
&\ds
{66721\, i\over 1512}\l_4^2\l_5^4 - 
{6301\over 90}\l_1\l_4^2\l_5^4 +
{50921 i\over 504}\l_1^2\l_4^2\l_5^4 - 
{2045\over 84}\l_1^3\l_4^2\l_5^4 + 
{1474\over 21}\l_3\l_4^2\l_5^4 - \quad\quad\quad &
\nonumber\\
&\ds
\quad\quad\;\; {19549\, i\over 72}\l_1\l_3\l_4^2\l_5^4 -
{75367\over 168}\l_1^2\l_3\l_4^2\l_5^4 + 
{21131\, i\over 63}\l_1^3\l_3\l_4^2\l_5^4 + 
{10258\over 135}\l_4^3\l_5^4 - 
{172369 \, i\over 1512}\l_1\l_4^3\l_5^4 - &
\nonumber\\
& \ds
\quad\quad\;\;\;
{87209\over 504}\l_1^2\l_4^3\l_5^4 + 
{21131\, i\over 189}\l_1^3\l_4^3\l_5^4 - 
{80497\, i\over 504}\l_3\l_4^3\l_5^4 - 
{12256\over 27}\l_1\l_3\l_4^3\l_5^4 +
{72071\, i\over 126}\l_1^2\l_3\l_4^3\l_5^4 + &
\nonumber\\
&\ds
{42262\over 189}\l_1^3\l_3\l_4^3\l_5^4 - 
{161701\over 1512}\l_3^2\l_4^3\l_5^4 + 
{29809\, i\over 126} \l_1 \l_3^2 \l_4^3 \l_5^4 +
{29809\over 126}\l_1^2\l_3^2\l_4^3\l_5^4 \quad\quad\quad\quad &
\label{P15}
\eeq

\beq
P_2^{(5)} \quad = 
& \ds 
-{41011\over 15120} + 
{202861\, i\over 7560}\l_3 + 
{3523\over 90}\l_1\l_3 + 
{6121\over 126}\l_3^2 - 
{3433\, i\over 36}\l_1\l_3^2 - 
{99\over 4}\l_1^2\l_3^2 - 
{86029\, i\over 1890}\l_3^3 - \quad &
\nonumber\\
&\ds
{4813\over 90}\l_1\l_3^3 +  
{130 i\over 9}\l_1^2\l_3^3 + 
{40\over 9}\l_1^3\l_3^3 -
{40799\, i\over 1890}\l_5 - 
{596233\over 5040}\l_3\l_5 + 
{5315\, i\over 24}\l_1\l_3\l_5 +\quad\quad &
\nonumber\\
&\ds
\quad\quad\quad {582137\, i\over 5040}\l_3^2\l_5 +
{76147\over 120}\l_1\l_3^2\l_5 -
{1640\, i\over 3}\l_1^2\l_3^2\l_5 + 
{350\over 9}\l_3^3\l_5 - 
{16229 i\over 36}\l_1\l_3^3\l_5 - 
%\nonumber\\
%&\ds
{2825\over 4}\l_1^2\l_3^3\l_5 + &
\nonumber\\
&\ds
{1660\, i\over 9}\l_1^3\l_3^3\l_5 - 
{2993\over 56}\l_5^2 +
%\nonumber\\
%&\ds
{1468283\, i\over 5040}\l_3\l_5^2 + 
{9213\over 20}\l_1\l_3\l_5^2 +
{35219\over 112}\l_3^2\l_5^2 - 
%\nonumber\\
%& \ds  
{28679\, i\over 24}\l_1\l_3^2\l_5^2 -\quad &
\nonumber\\
&\ds
{6715\over 8}\l_1^2\l_3^2\l_5^2 - 
{15121\, i\over 126}\l_3^3\l_5^2 -
%\nonumber\\
%&\ds
{23695\over 36}\l_1\l_3^3\l_5^2 + 
{3115\, i\over 4}\l_1^2\l_3^3\l_5^2 + 
{940\over 9}\l_1^3\l_3^3\l_5^2 + 
{47321\, i\over 1260}\l_5^3 + \quad &
\nonumber\\
&\ds
{209881\over 1080}\l_3\l_5^3 - 
{9325\, i\over 36}\l_1\l_3\l_5^3 -
%\nonumber\\
%&\ds
{81509\, i\over 504}\l_3^2\l_5^3 -
{15457\over 36}\l_1\l_3^2\l_5^3 + 
{340\, i\over 3}\l_1^2\l_3^2\l_5^3 - 
{4810\over 189}\l_3^3\l_5^3 + \quad &
\nonumber\\
&\ds
{1801\, i\over 18}\l_1\l_3^3\l_5^3 - 
{3055\over 18}\l_1^2\l_3^3\l_5^3 + 
160\, i \l_1^3\l_3^3\l_5^3 + 
%\nonumber\\
%&\ds
{6599\over 378}\l_5^4 -
{1151\, i\over 21}\l_3\l_5^4 + 
{53\over 24}\l_1\l_3\l_5^4 -  
%\nonumber\\
%&\ds
{1999\over 126}\l_3^2\l_5^4 - &
\nonumber\\
&\ds
{599\, i\over 4}\l_1\l_3^2\l_5^4 - 
{4225\over 24}\l_1^2\l_3^2\l_5^4 -
{113\, i\over 9}\l_3^3\l_5^4 - 
%\nonumber\\
%&\ds
{946\over 9}\l_1\l_3^3\l_5^4 + 
240\, i \l_1^2 \l_3^3 \l_5^4 +
80\l_1^3\l_3^3\l_5^4\quad\quad\quad\quad\;\; & 
\nonumber\\
\phantom{a} 
&\quad & 
\label{P25} 
\eeq

\end{document}